\documentclass[aps,pra,amsmath,amssymb,floatfix,nofootinbib,twocolumn,superscriptaddress]{revtex4-2}

\usepackage[T1]{fontenc}
\usepackage[T1]{tipa}
\usepackage{graphicx}
\usepackage{physics}
\usepackage{bm}
\usepackage{times}
\usepackage{natbib}
\usepackage{jabbrv}

\begin{document}

\title{Compact Fiber-Coupled Narrowband Two-Mode Squeezed Light Source}

\author{Umang Jain}
\affiliation{Homer L. Dodge Department of Physics and Astronomy, University of Oklahoma, Norman, OK, 73019, USA}
\affiliation{Center for Quantum Research and Technology (CQRT), University of Oklahoma, Norman, OK, 73019, USA}

\author{Jae Choi}
\altaffiliation{Currently at Opto-Atomics Corp., San Pedro, CA 90731, USA}
\affiliation{Hedgefog Research Inc., San Pedro, CA 90731, USA}

\author{Christopher Hull}
\altaffiliation{Currently at EOSPACE Inc., Redmond, WA 98052, USA}
\affiliation{Hedgefog Research Inc., San Pedro, CA 90731, USA}

\author{Alberto M. Marino}
\email{marino@ou.edu, marinoa@ornl.gov}
\affiliation{Homer L. Dodge Department of Physics and Astronomy, University of Oklahoma, Norman, OK, 73019, USA}
\affiliation{Center for Quantum Research and Technology (CQRT), University of Oklahoma, Norman, OK, 73019, USA}
\affiliation{Quantum Information Science Section, Computational Sciences and Engineering Division, Oak Ridge National Laboratory, Oak Ridge, TN, 37831, USA}
\thanks{This manuscript has been authored in part by UT-Battelle, LLC, under contract DE-AC05-00OR22725 with the US Department of Energy (DOE). The publisher acknowledges the US government license to provide public access under the DOE Public Access Plan (http://energy.gov/downloads/doe-public-access-plan)}

\begin{abstract}
Quantum correlated states of light, such as squeezed states, are a fundamental resource for the development of quantum  technologies, as they are needed for applications in quantum metrology, quantum computation, and quantum communications. It is thus critical to develop compact, efficient, and robust sources to generate such states. Here we report on a compact, narrowband, fiber-coupled source of two-mode squeezed states of light at 795~nm based on four wave mixing (FWM) in a $^{85}$Rb atomic vapor. The source is designed in a small modular form factor, with two input fiber-coupled beams, the seed and pump beams required for the FWM, and two output fibers, one for each of the modes of the squeezed state. The system is optimized for low pump power (135~mW) to achieve a maximum intensity-difference squeezing of 4.4~dB after the output fibers at an analysis frequency of 1~MHz. The narrowband nature of the source makes it ideal for atomic-based quantum sensing and quantum networking configurations that rely on atomic quantum memories. Such a source paves the way for a versatile and portable platform for applications in quantum information science.
\end{abstract}

\maketitle

\section{Introduction}

Squeezed light has become an indispensable resource for quantum sensing, quantum communications, and quantum computing~\cite{Andersen_2016, Lawrie2019, Polzik1992, PhysRevApplied.14.034065, Xu2019, Andersen2015, Pirandola2018, Weedbrook2012}. Its low noise properties, as compared to classical coherent states, makes it possible to increase the sensitivity and precision of a measurement~\cite{Lawrie2019}. For example, squeezed states of light have been used to enhance the sensitivity of the Laser Interferometer Gravitational-wave Observatory (LIGO), leading to significant advances for gravitational wave detection~\cite{2011, Aasi2013}. They also serve as a resource for continuous-variable quantum computing~\cite{PhysRevLett.97.110501, PhysRevLett.82.1784}, high-fidelity quantum teleportation~\cite{PhysRevLett.80.869}, and secure quantum communication protocols~\cite{PhysRevLett.88.057902,usenko2025}.

In order to make squeezed light more accessible for real-world applications, it is necessary to develop sources that are modular, robust, and that offer low size, weight, and power (SWaP). Implementing these sources such that the squeezed light is fiber coupled makes them more robust for field deployable applications or when operating in harsh environments, such as at cryogenic temperatures, and facilitates delivery to the required apparatus.
There have been efforts to develop modular squeezed light sources with optical parametric oscillators~\cite{Arnbak2019}, parametric down conversion in nonlinear crystal waveguides, both free space~\cite{Kashiwazaki2020} and fiber coupled~\cite{9044771}, and FWM in optical fibers~\cite{Liu_2024}; however, there hasn't been a significant effort to develop fiber-coupled narrowband squeezed light sources. Such narrowband sources are critical for applications involving atomic systems, such as quantum memories~\cite{PhysRevLett.100.093602, RN966}, quantum sensors~\cite{5778937,PhysRevLett.132.190001}, and atomic clocks~\cite{Colombo2022, Schulte2020}, as the quantum states of light need to have a bandwidth that matches that of the involved atomic transitions to obtain an efficient interaction. Given the push for deployable quantum technologies, many of which are atomic based, it is important to develop low SWaP fiber-coupled narrowband sources of quantum states of light.

We present a compact fiber-coupled source of narrowband two-mode squeezed states of light generated through a FWM process in a $^{85}$Rb atomic vapor cell. While there have been previous efforts that use this process to implement compact systems based on diode lasers and acousto-optic modulators (AOMs) or fiber coupled electro-optic modulators (EOMs)~\cite{Qin.12,PhysRevLett.113.023602,PhysRevA.96.043843,Sim.25}, they focused on the generation of the input beams required for the FWM. In contrast, our system is optimized for compactness, low optical power requirements, and coupling of the quantum states to optical fibers to deliver over 4~dB of intensity-difference squeezing at the output of single mode fibers.

\section{Experimental Setup}

\begin{figure}[ht]
\centering
\includegraphics[width=\linewidth]{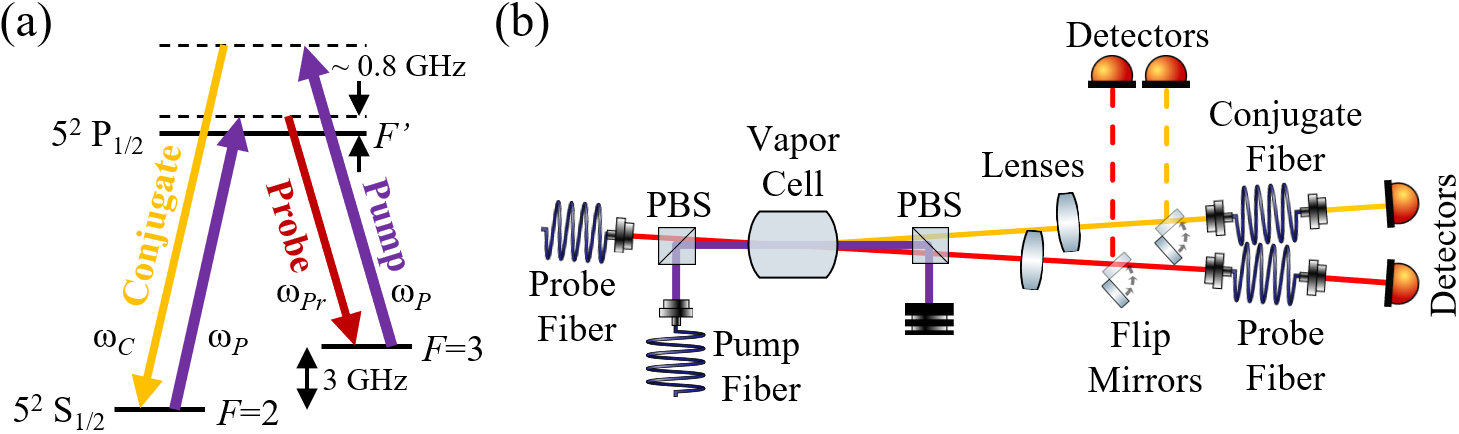}
\caption{(a) Double-$\Lambda$ configuration in $^{85}$Rb used for the generation of twin beams.  As a result of the FWM process, two pump photons (purple) are absorbed and two photons, which we call probe (red) and conjugate (yellow), are emitted. The frequencies of the pump ($\omega_{P}$), probe ($\omega_{Pr}$), and conjugate ($\omega_{C}$) are such that the involved fields conserve energy. (b) Schematic of the experimental setup for the generation of  fiber-coupled twin beams through a FWM process in a hot Rb vapor cell.}
\label{fig:double_lambda}
\end{figure}

We use a FWM process based on a double-$\Lambda$ configuration in hot $^{85}$Rb vapor (see Fig.~\ref{fig:double_lambda}(a) for energy level scheme) to generate narrowband two-mode squeezed states of light or twin beams at 795~nm~\cite{PhysRevA.78.043816}. In this process, two pump photons (at frequency $\omega_{p}$) and one seed photon at the probe frequency ($\omega_{pr}$) interact with the atomic vapor to emit photons in pairs, one at the probe frequency and a second one at the conjugate frequency ($\omega_{c}$).  A Ti:Sapphire laser at 795~nm is used to generate the required strong pump beam. A small portion of the laser is downshifted by 3.04~GHz (energy separation between the \textit{F}=2 and \textit{F}=3 hyperfine energy levels of the $5^{2}S_{1/2}$ ground state) with a high frequency AOM to generate the probe beam, such that the pump and probe are close to the two-photon resonance between the ground state hyperfine levels.

As a result of conservation of energy, which leads to the simultaneous generation of probe and conjugate photons, the probe beam is amplified and a conjugate beam is generated. Such simultaneous emission of photons leads to temporal quantum correlations between the amplitudes of the probe and conjugate beams and makes it possible to obtain noise levels below the shot noise limit (SNL). These correlations can be characterized by measuring the powers of the probe and conjugate with photodetectors, obtaining the difference between the generated photocurrents, and using a spectrum analyzer to determine the intensity-difference noise.

Figure~\ref{fig:double_lambda}(b) shows the schematic of the experimental setup. Implementation of the FWM requires orthogonally polarized probe and pump beams overlapping at a slight angle at the center of a Rb vapor cell.  To implement such a configuration, two independent single mode, polarization maintaining optical fibers are used to deliver the orthogonally polarized probe and pump beams. After the fibers, pump and seed probe beams are made to intersect at an angle of 0.4$^\circ$ at the center of a 12~mm long (12~mm diameter) isotopically pure $^{85}$Rb vapor cell that is heated to $\sim 99^{\circ}$C. After the FWM process, the twin beams are coupled into two  single mode, polarization maintaining optical fibers with AR coated inputs, after which intensity-difference squeezing measurements are performed.  Flip mirrors are placed before the output fibers to bypass them in order to obtain a direct measure of the squeezing level generated by the FWM process.

In our previous experiments \cite{Dowran2018} we have shown that the FWM process is capable of generating twin beams with 9~dB of intensity-difference squeezing. This was achieved when using a pump beam with an optical power of 550~mW and a size of 1~mm $1/e^{2}$ waist diameter and an input seed probe that was downshifted by 3.04~GHz and had a size of 0.7~mm $1/e^{2}$ waist diameter. For the fiber-coupled twin beam source, these parameters are optimized such that maximum amount of squeezing is preserved after fiber coupling and the pump power can be kept as low as possible for a low power and robust setup.

Given that the generated twin beams are spatially multimode~\cite{Boyer08, Corzo2011, PhysRevA.93.063821}, care has to be taken when coupling them to single mode optical fibers.  It is necessary to couple, as much as possible, only the correlated spatial regions or modes of the twin beams. For a seeded FWM process, as the one used here, all the modes supported by the process will be generated, with the ones that have a spatial overlap with the input seed generated as bright twin beams and the other ones as vacuum twin beams. Thus, ideally it is possible to couple only correlated modes by perfectly  fiber coupling (perfect spatial mode matching to the fiber mode) the bright portions of the twin beams. However, for an imperfect spatial mode matching portions of the generated vacuum twin beams will also couple into the fiber. Once inside the fiber, all coupled modes will be in the spatial mode defined by the fiber and thus will contribute to the measured noise. In particular, coupling of uncorrelated probe/conjugate spatial regions or modes will lead to excess noise given that each beam by itself has thermal statistics, and hence will lower the squeezing. This is analogous to the excess noise coupled in homodyne detection when the twin beams are not properly mode matched to their corresponding local oscillators~\cite{Gupta2020}.

\begin{figure}[ht]
\centering
\includegraphics[width=0.9\linewidth]{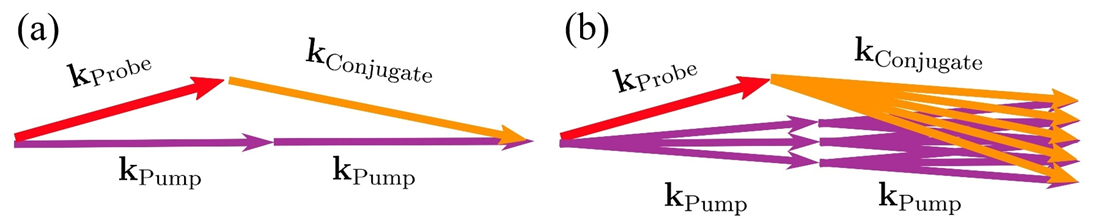}
\caption{Momentum correlations due to phase matching condition. (a) A pump beam with a single $k$-vector (infinite plane wave) will lead to correlations between a single probe $k$-vector and a single conjugate $k$-vector. (b) A pump with a finite spatial extent will have a spread of $k$-vectors, which will lead to correlations between a given probe $k$-vector and a distribution of conjugate $k$-vectors.}
\label{fig:Phase Matching}
\end{figure}

The spatially multimode nature of the twin beams, which result in position-momentum quantum correlations between them~\cite{Kumar_2021}, are dictated by the phase matching condition, as shown in Fig.~\ref{fig:Phase Matching}. For the ideal case of an infinite plane-wave pump, for which the angular spectrum of the pump is a single $k$-vector, the FWM would generate a conjugate beam with a single $k$-vector for a given $k$-vector of the probe, as shown in Fig.~\ref{fig:Phase Matching}(a). This means that in the far field one point in the transverse profile of the probe field is only correlated to one point in the transverse profile of the conjugate field, i.e. point-to-point correlations. On the other hand, for the realistic case of a finite pump beam size, for which the pump has an angular spectrum with a spread of $k$-vectors, a given $k$-vector of the probe will be correlated with a distribution of $k$-vectors for the conjugate, as shown in Fig.~\ref{fig:Phase Matching}(b). In the far field, this leads to probe-conjugate correlations between subregions of the transverse spatial profiles of the beams. The smallest size of such independently correlated subregions is known as the coherence area \cite{PhysRevA.93.063821,PhysRevA.98.043853}.

To mainly couple correlated coherence areas of the probe and conjugate beams into the fibers, we match as much as possible the size of the bright portion of the twin beams to that of a single coherence area and optimize the fiber coupling of the bright regions. To this end, we make the size of the pump and input probe seed equal to each other, with $1/e^{2}$ waist diameters, at the center of the cell, of $\approx0.6$~mm. Making the size of the pump small has the advantage that it reduces the overall number of spatial modes that the system can support (larger coherence area)~\cite{PhysRevA.93.063821}, which makes the setup more robust to changes in alignment and drift, and reduces the amount of total optical power needed for the FWM. With this configuration we obtain maximum squeezing for pump powers in the range of $\sim 135$~mW to $\sim 250$~mW, with powers outside this range leading to reduced gain or increased noise. Given the aim of a low SWaP system, a pump power of $\sim 135$~mW is used.

\begin{figure}[ht]
\centering
\includegraphics[width=0.85\linewidth]{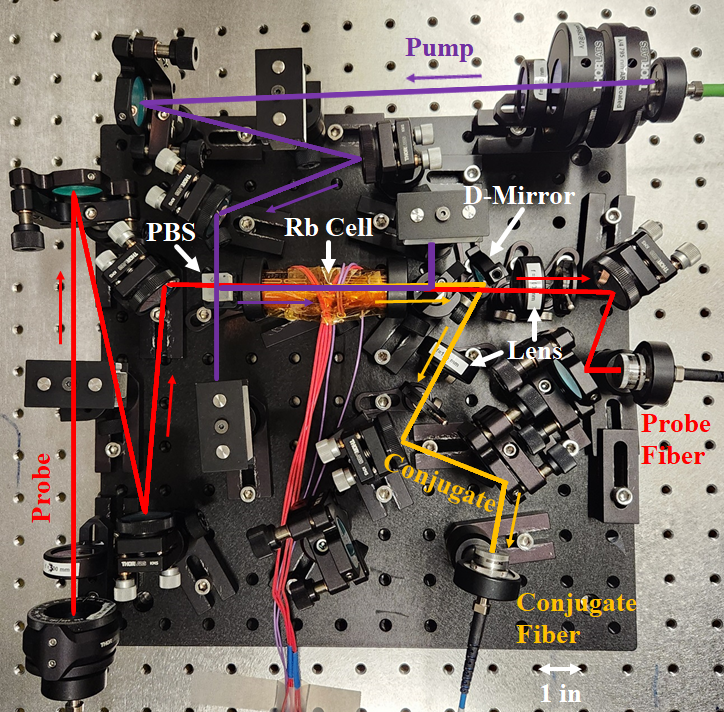}
\caption{Picture of system implemented on a 12 in by 12 in breadboard. The overlaid lines show the paths of the pump beam (purple), the probe beam (red), and the conjugate (yellow). The scale bar (white double arrow at the bottom of picture) corresponds to a length of 1 in.}
\label{fig:Real Setup}
\end{figure}

Figure~\ref{fig:Real Setup} shows a picture of the setup used to generate the fiber-coupled twin beams. The amplified probe and generated conjugate beams are each coupled to dedicated single mode, polarization maintaining fiber using an $f$-to-$f$ optical system implemented with a 100~mm focal length lens for the probe beam and 125~mm for the conjugate beam. Different focal lengths are required as the cross-Kerr effect, due to a strong pump, affects the spatial profile of the probe more than the conjugate given that it is significantly closer to atomic resonance.  This leads to slightly different sized probe and conjugate beams. The $f$-to-$f$ configuration ensures that the fiber input couplers for both probe and conjugate  are placed at the fourier plane of the center of the cell, i.e. far field. With this configuration we are able to keep the setup compact while obtaining fiber coupling efficiencies greater than 90\% for both probe and conjugate beams.

To characterize the level of squeezing, we calibrate the SNL with a classical (coherent) state with the same power as the total combined power of the probe and conjugate. To obtain an accurate calibration, we use the probe beam with the pump beam blocked, such that there is no FWM. A half-wave plate and a polarizing beamsplitter are then used to split the beam into two beams of equal power that are sent to a balanced detector to perform intensity-difference noise measurements. This makes it possible to cancel any classical technical noises and obtain an accurate measure of the SNL.

\section{Results}

\begin{figure}[ht]
\centering
\includegraphics[width=0.85\linewidth]{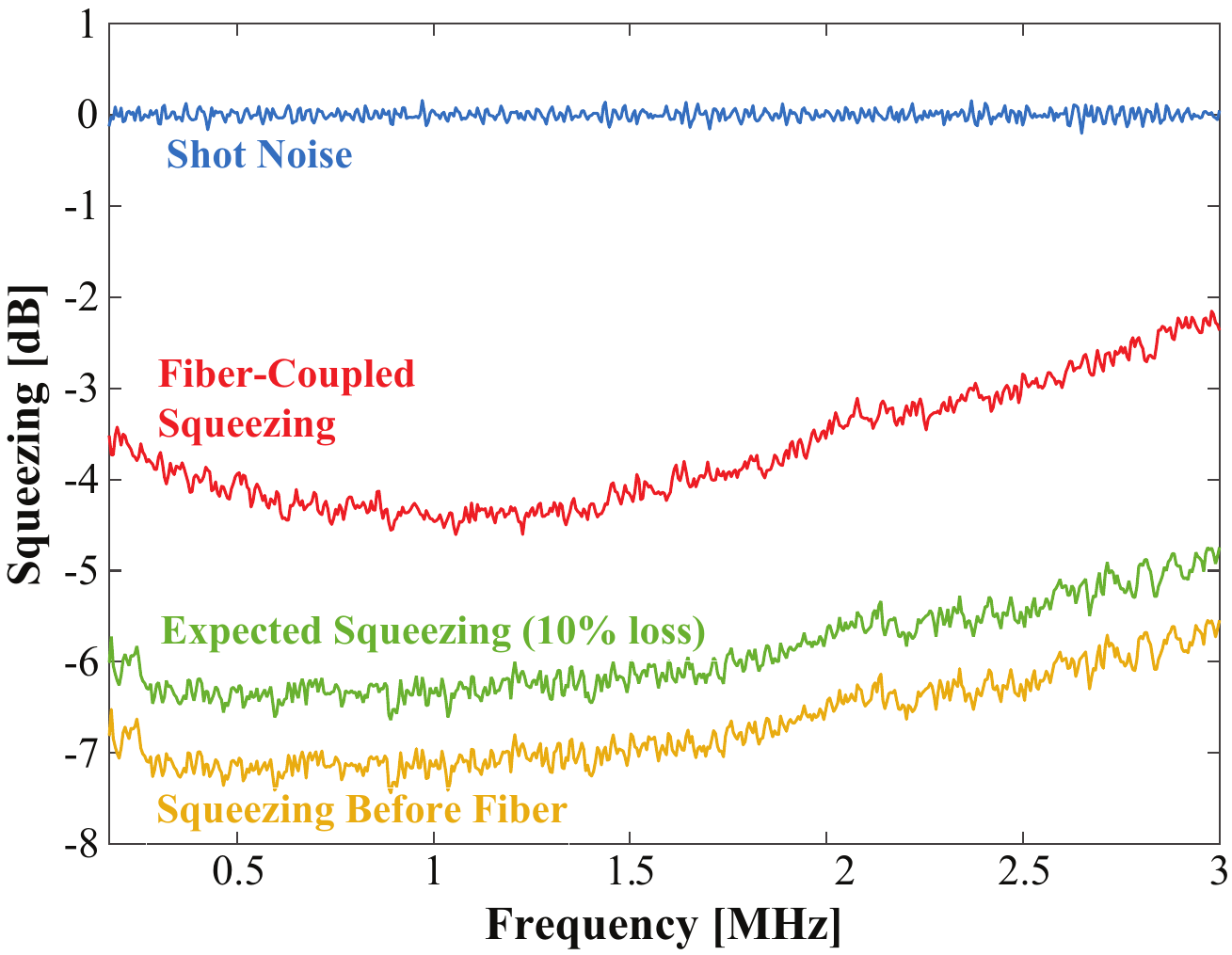}
\caption{Intensity-difference noise measurements normalized to the shot noise limit (blue trace) for the twin beams before (yellow trace) and after (red trace) the output optical fibers. The green trace shows the expected level of squeezing if we assume that the 10\% fiber coupling inefficiency is due only to optical losses. The power of the generated bright squeezed beams is $\sim 200~\mu$W per beam.}
\label{fig:Result Plot}
\end{figure}

The measured intensity-difference noise levels normalized to the SNL are shown in logarithmic scale in Fig.~\ref{fig:Result Plot}, such that the SNL (blue trace) is at 0~dB. To characterize the impact of fiber coupling the twin beams, we compare the squeezing level that is measured  at the output of the optical fiber to the one right after the vapor cell. To this end, flipped mirrors are used to bypass the output fibers and send the twin beams directly to the balanced detector. When we bypass the optical fibers, we measure an intensity-difference squeezing level of $-7.2$~dB in the frequency range from 0.3~MHz to 1~MHz, as shown by the yellow trace in Fig.~\ref{fig:Result Plot}, with the level of squeezing decreasing at higher frequencies due to the narrow bandwidth of the FWM process. This measurement provides an initial characterization of the level of quantum correlations produced by the source.

With $> 90$\% coupling efficiency of the probe and conjugate beams into their corresponding optical fiber, we obtain $-4.4$~dB of intensity-difference squeezing at the output of the fibers, as shown by the red trace in Fig.~\ref{fig:Result Plot}. We also compare the measured level of squeezing with the level we would expect with 10\% optical losses for both probe and conjugate. As shown by the green trace in Fig.~\ref{fig:Result Plot}, this level of loss should lead to a reduction of the level of squeezing from -7.2~dB to -6.3~dB at  1~MHz.

The discrepancy of $\sim 1.9$~dB between the expected and measured levels of squeezing when taking into account 10\% optical losses is mainly due to the coupling of uncorrelated subregions of the twin beams along with the correlated ones. Two factors in particular can lead to this: imperfect mode
matching of corresponding spatial modes into the fundamental mode of the fibers and not coupling the probe and conjugate into their corresponding fiber at exactly the same optical plane (far field for our experiments), both of which are complicated due to the cross-Kerr effect introduced by the FWM.

To quantify the impact of coupling uncorrelated spatial modes into the fiber, we consider a model in which the fiber coupling inefficiency is treated as a beamsplitter with both vacuum noise (from regular optical losses) and thermal noise (due to uncorrelated modes) coupling into the system through the unused beamsplitter port, as was done in~\cite{Gupta2020}. For the case in which the fiber coupling efficiency ($\eta$) is the same for both probe and conjugate and all the twin beam spatial modes that are coupled into the fiber have the same squeezing parameter, the expected level of squeezing ($S'$) is given by  $S'=\eta S_0+(1-\eta)[\epsilon N_{\rm th}+(1-\epsilon) N_{\rm v}]$, where $S_0$ is the initial level of squeezing, $N_{\rm th}$ is the normalized thermal noise contribution of coupled uncorrelated spatial modes, $N_{\rm v}=1$ is the normalized vacuum noise, and $\epsilon$ and $(1-\epsilon)$ are the fractions of thermal noise and vacuum noise that couple through the unused port of the beamsplitter, respectively, with $\epsilon=1$ corresponding to only thermal noise coupling through the unused beamsplitter port. For a coupling efficiency of 90\%, initial squeezing of -7.2 dB, normalized excess thermal noise consistent with that of the initial level of squeezing, and equal fractions of coupled thermal and vacuum noise, the expected level of squeezing is -3.2 dB. While this simple model overestimates the fraction and the contribution of uncorrelated spatial modes to the total noise, it shows its impact on the level of squeezing.

Further optimization of the squeezing level after the output optical fibers requires a reduction in optical losses and optimization of the spatial mode matching into the fibers. One of the biggest contributions to optical losses is the reflections of the output ends of the fiber, which are not AR coated. To enhance the spatial mode matching, it would be necessary to optimize the spatial modes of the generated probe and conjugate, which can be done by modifying the spatial profile of the input seed probe directly or of the pump beam. As we have previously shown, changing the spatial profile of the pump directly impacts the spatial properties of the generated spatial modes of the twin beams~\cite{PhysRevA.93.063821,sciadv.adf9161}.

\section{Conclusion}

We have demonstrated a compact, fiber-coupled, narrowband source of two-mode squeezed states of light capable of delivering over 4~dB of intensity-difference squeezing. In addition to being fiber coupled, the optimized configuration uses a smaller pump beam diameter that makes it possible to operate at pump power levels significantly lower than previous experiments based on FWM in atomic vapors. Such a design demonstrates the possibility of implementing a low-SWaP source of squeezed light. The need for such a source stems from the growing demand for deployable quantum technologies for applications in sensing, networking, and precision measurements.  This is particularly relevant for space, defense, and mobile applications where SWaP constraints are critical. Thus, our compact design promises to make two-mode squeezed states of light more accessible for deployable quantum devices with applications towards quantum information science.

\section*{Acknowledgments}
This research was performed as part of a collaboration between Hedgefog Research Inc. and the University of Oklahoma under NASA Contract 80NSSC21C0359. Information contained in this publication may be subject to SBIR/STTR Data Rights per FAR 52.227-20.

\end{document}